\documentclass[a4paper,fleqn,usenatbib,letter]{mnras}

\usepackage{epsfig}
\usepackage{times}
\usepackage{amsmath}

\newif\ifAMStwofonts

\newcommand{\target}{V404\,Cyg}
\newcommand{\Msun}{$M_{\odot}$}

\newcommand{\sloanr}{$\it r'$}

\title[Evidence for magnetic field compression in shocks within the jet of V404\,Cyg]
{Evidence for magnetic field compression in shocks within the jet of V404\,Cyg}

\author[T.\,Shahbaz et al. ]
       {T.\,Shahbaz$^{1,2}$,
       \thanks{E-mail: tsh@iac.es (TS)}
	    D.M.\,Russell$^{3}$, 
	    S.\,Covino$^{4}$,
	    K.\,Mooley$^{5}$,		
	    R.P.\,Fender$^{5}$ and		
	    C.\,Rumsey$^{6}$ 		\\
$^{\it 1}$Instituto de Astrof\'\i{}sica de Canarias (IAC), E-38200 La Laguna, 
Tenerife, Spain \\
$^{\it 2}$Dept. Astrof\'\i{}sica Universidad de La Laguna (ULL), E-38206 
La Laguna, Tenerife, Spain \\
$^{\it 3}$New York University Abu Dhabi, PO Box 129188, Abu Dhabi, United Arab 
Emirates \\
$^{\it 4}$INAF, Osservatorio Astronomico di Brera, Via E. Bianchi 46, 
I-23807 Merate (Lc), Italy \\
$^{\it 5}$Astrophysics, Department of Physics, University of Oxford, 
Keble Road, Oxford OX1 3RH, UK \\
$^{\it 6}$Astrophysics Group, Cavendish Laboratory, 19 J. J. Thomson Avenue, 
Cambridge CB3 0HE, UK
}

\pagerange{\pageref{firstpage}--\pageref{lastpage}}
\pubyear{2014}

\begin{document} 
\maketitle 
\begin{abstract} 

\noindent
We present optical and near-IR linear polarimetry of \target\  during its 2015
outburst and in quiescence. We obtained time resolved $r'$-band polarimetry 
when the source was in outburst, near-IR polarimetry when the source  was near
quiescence and multiple wave-band optical polarimetry later in quiescence.   The
optical to near-IR linear polarization spectrum can be described by 
interstellar  dust and an intrinsic variable component. The intrinsic optical
polarization, detected during the rise of one of the brightest flares of the
outburst,  is variable, peaking at 4.5 per cent and decaying to 3.5 per cent. We
present several arguments that favour a synchrotron jet origin to this variable
polarization, with the optical emission originating close to the jet base.  The
polarization flare occurs during the initial rise of a major radio flare event
that peaks later, and is consistent with a classically evolving synchrotron flare
from an ejection event. We conclude that the optical polarization flare represents a
jet launching event; the birth of a major ejection.  For this event we measure a
rather stable polarization position angle of  -9$^\circ$ E of N, implying that
the magnetic field near the base of the jet is approximately perpendicular to
the jet axis. This may be due to the compression of magnetic field lines in
shocks in the accelerated plasma, resulting in a partially ordered transverse
field that have now been seen during the 2015
outburst. 
We also find that this ejection occurred at a similar stage in the
repetitive cycles of flares.

\end{abstract} \begin{keywords} 
binaries: close -- stars: fundamental parameters -- stars: individual: V404\,Cyg  --  
stars: neutron -- 
X-rays: binaries
\end{keywords}

\section{INTRODUCTION}

Polarimetry can provide a unique probe of X-ray binaries, since (1) we can obtain
information on the geometrical property of the binary system from the light 
scattered by circumstellar matter (stellar wind and/or an accretion disc) at  a
scale not accessible to present imaging techniques or (2) information  on
jets (e.g., geometry, magnetic field, electron energy) from the  synchrotron
emission. A significant fraction of the optical emission in 
low-mass X-ray binaries (LMXBs) is
produced by the accretion disc surrounding the compact object. The  accretion
disc is heated by X-rays originating near the central source  and hydrogen is 
completely ionized in the entire disc
\citep{vanParadijs94,Charles06,Spruit14}.

Therefore,  optical linear polarization at
the few per cent level caused by Thomson scattering of the emitted radiation 
with the free electrons in the  disc is expected  at these wavelengths
\citep{Brown78,
Dolan84, Cheng88, Schultz00}. The polarization  due to the scattering of
intrinsically unpolarized thermal emission can be  modulated on the orbital
period, which places constraints on the physical  and geometrical properties of
the system \citep{Dolan89, Gliozzi98}.

In LMXBs, there is also one emission mechanism 
that intrinsically produces polarized light -- synchrotron emission from 
particles in a sufficiently well-ordered magnetic field in a collimated 
outflow. Optically thin synchrotron emission is intrinsically polarized. 
If the local magnetic field in the emitting region is uniform (ordered), a 
net linear polarization is observed. If the field is tangled, the 
differing angles of the polarized light suppress the observed, average 
polarization. In the case of a perfectly ordered field the maximum 
polarization strength is $\sim$70--80 percent, 
\citep{Rybicki79,Bjornsson82} and is dependent only on the degree of 
ordering of the field and the energy distribution of the electron 
population. The polarimetric measurements of the optically thin 
synchrotron emission provide a powerful tool to probe the nature of 
magnetic field structure.

\target\ (=GS\,2023+338), was discovered with the all-sky monitor aboard 
Ginga on 1989 May 22 \citep{Makino89}.  Previously, \target\ had  
at least two outbursts, in 1938 and 1956, which were only observed at 
optical wavelengths because of the lack of X-ray space instruments at that time.
This lack of X-ray
knowledge meant that \target\ was mis-classified as a classical (albeit
recurrent) nova.  

\target\ is known to contain a K0\,III-V secondary star orbiting a 
$\sim$10\Msun black hole \citep{Shahbaz94,Shahbaz96} with an orbital period of 
6.5\,d \citep{Casares92} and is located at a precise distance of 2.4\,kpc, 
obtained using radio parallax measurements \citep{Miller-Jones09}. During 
outburst \target\ varies in X-ray flux by a factor of $\sim$500 on 
timescales of seconds \citep{Kitamoto89} and a factor of 2--10 on 
timescales of 30\,min to a few hours in quiescence \citep{Wagner94, 
Kong02, Hynes04}.

Recently, on 2015 June 15 (MJD\,57188), \target\ went into its fourth
outburst. It was first detected by the X-ray satellite Swift 
\citep[BAT and XRT;][]{Barthelmy15} and then with MAXI, and INTEGRAL 
\citep{Negoro15,Kuulkers15}. The outburst, triggered by a 
viscous-thermal instability close to the inner 
edge of the truncated accretion disc \citep{Bernardini16}. The early 
ATEL alerts triggered follow-up observations at all wavelengths, from radio 
\citep{Mooley15} to hard X-rays and the source reached up to 50 Crab in 
hard X-rays \citep{Rodriguez15}, with extreme 
flaring activity at all wavelengths \citep[e.g.][]{Mooley15,Garner15, 
Ferrigno15, Motta15a,Motta15b,Natalucci15,Tetarenko15, Tsubono15, 
Kimura16, Gandhi16, Marti16}. In this paper we present the results of our 
optical and near-IR polarimetry of \target\ during its 2015 outburst and 
in quiescence after the outburst.

\begin{table}
\caption{Results of the optical and near-IR  polarimetric observations 
of \target\ in, and near to, quiescence.}
\begin{center}
\begin{tabular}{llcc}\hline 
\textsc{ut} Date & Band        & $p$ (\%)      & $\theta$ ($^{\circ}$)  \\ \hline
2016 May 26 & $V$          &  7.69$\pm$0.61  &    5.1 $\pm$2.1 \\
2016 May 26 & $R$          &  7.41$\pm$0.32  &    7.2 $\pm$1.1 \\
2016 May 26 & $I$          &  6.00$\pm$0.16  &    6.3 $\pm$0.7 \\
2016 May 26 & $z$          &  5.36$\pm$0.14  &    6.9 $\pm$0.6\\
2015 July 27 & $J$          &  3.33$\pm$0.05  &   -7.5 $\pm$4.3 \\
2015 July 27 & $H$          &  2.22$\pm$0.04  &   -9.9 $\pm$3.8 \\
2015 July 27 & $K_{\rm s}$  &  1.33$\pm$0.08  &    0.2 $\pm$12  \\
\hline       
\end{tabular}
\end{center}
\label{table:results}
\end{table}

\section{OBSERVATIONS AND DATA REDUCTION}

\subsection{TNG optical observations}
\label{sec:optical}

Time-resolved optical polarimetry of \target\ was obtained with the 3.5\,m 
Telescopio Nazionale Galileo (TNG) at the Observatorio del Roque de los 
Muchachos, La Palma, Spain, during the night of \textsc{ut} 2015 June 23. 
The polarimeter PAOLO 
\citep[Polarimetric Add-On for the LRS Optics;][]{Covino14} was used, 
which consists of a double Wollaston prism mounted in the filter wheel, 
producing four simultaneous polarization states of the field of view.  
The optical setup of the polarimeter is  described in \citet{Oliva97}.
The images are separated by the prism producing four simultaneous 
image slices on the CCD that correspond to four different position angles 
with respect to the telescope axis (0$^\circ$, 45$^\circ$, 90$^\circ$ and 
135$^\circ$).  Measurements of the flux in these four images allows one to 
measure the degree of linear polarization of a source. Since the 
instrument is mounted on the Nasmyth focus of the TNG, the instrumental 
polarization of the order of 2--3 per cent varies with parallactic angle 
\citep{Giro03}. Therefore, in order to determine the instrumental 
polarization model we observed the zero polarized standard star HD\,154892 
over a wide range of parallactic angles. The \sloanr\ filter was used for 
all the observations and given the brightness of \target, we used an 
exposure time of 2\,s and obtained time-resolved polarimetry for 1.92\,hr 
(\textsc{ut} 2015-06-24 02:53 to 04:48; MJD\,57197.1211 to 57197.2003). 
Bias images were taken as well as dome flat-field images. The conditions 
were good with a median seeing of 0.9 arcsec.

\begin{figure}
\centering
\hspace*{-5mm}
\includegraphics[angle=-90,width=8.cm]{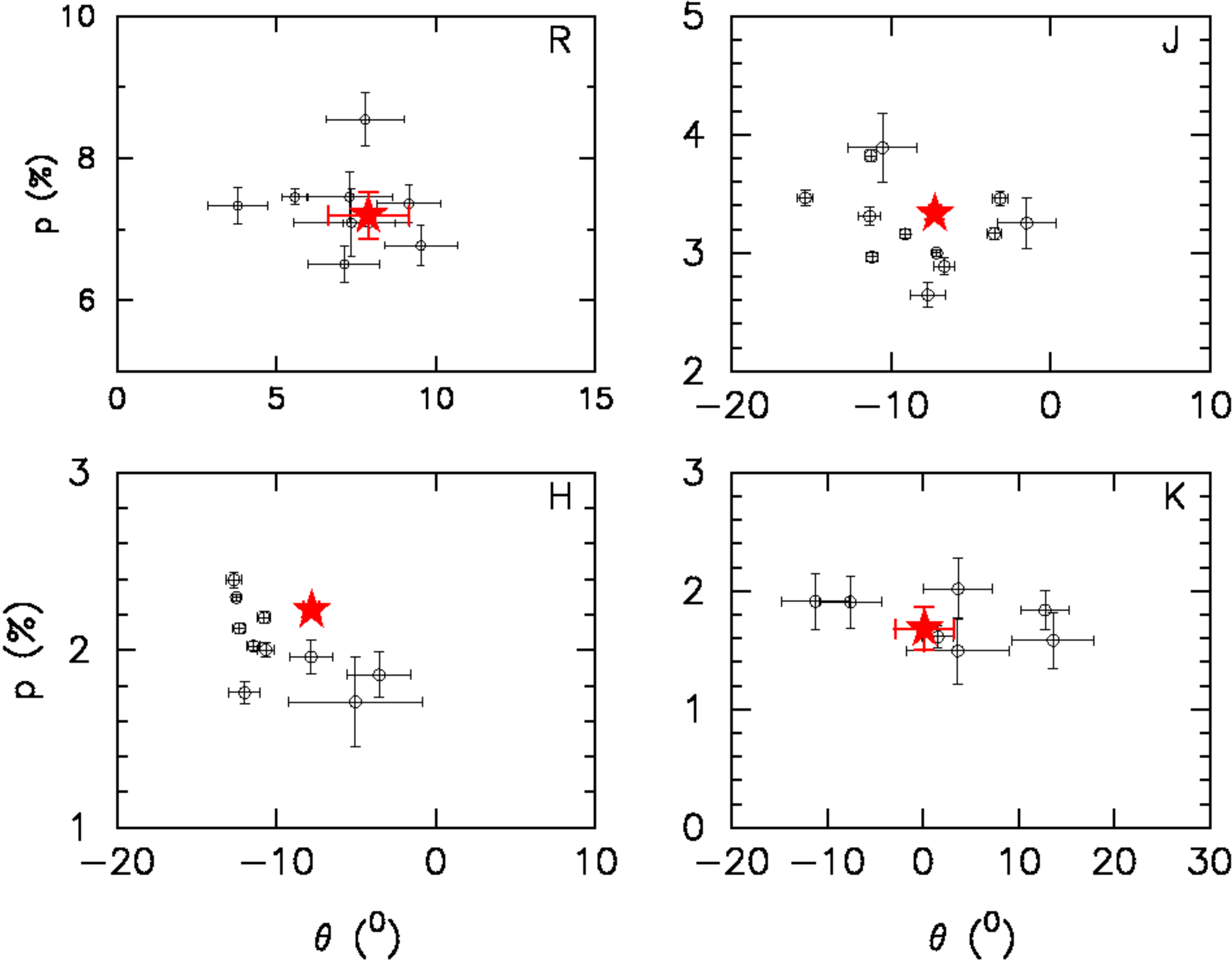}
\caption{
Optical (2016 May 24) and near-IR $JHK_{\rm s}$ (2015 July 27)   polarization
$p$ versus  position angle $\theta$ for \target\ (large filled red star) and
stars in the field of view  (circles).
}
\label{fig:ppa}
\end{figure}

\subsection{NOT optical observations}

Broad-band optical polarimetry of \target\ was obtained with the 2.5\,m 
Nordic Optical Telescope (NOT) at the Observatorio del Roque de los 
Muchachos, La Palma, Spain, during \textsc{ut} 2016 May 26 
($\sim$MJD\,57534.2). Linear polarimetry was performed using a half wave 
plate in the FAPOL unit and a calcite plate mounted in the aperture wheel. 
The calcite plate provides the simultaneous measurement of the ordinary 
and the extraordinary components of two orthogonally polarized beams. 
Images were taken at half wave plate rotation angle 0$^\circ$, 
22.5$^\circ$, 45$^\circ$, and 67.5$^\circ$ and the linear polarization 
($p$) and the position angle ($\theta$) were determined. We also observed 
a non-polarized star and polarized star to determine the instrumental and 
position angle offset, respectively. We obtained images in the $V$, $R$, $I$ 
and $z'$ bandpasses using exposure times of 300\,sec, 120\,sec, 120\,sec 
and 300\,sec, respectively. Bias images and dome flat-fields were taken. 
The conditions were good with a seeing of 0.7 arcsec.

\subsection{Optical data reduction}

The data reduction for the TNG and NOT data was performed using the 
\textsc{ultracam} software \citep{Dhillon07}. The flat-field images were 
first de-biased using the bias image and then combined to create a master 
flat-field. The science images were also de-biased and then flat-fielded 
in the standard way. Aperture photometry was then performed on the science 
targets, comparison stars and standard stars using a fixed aperture. 
For the quiescent data, the good seeing allows  us to mask 
the line-of-sight contaminating star 1.4 arcsec north of \target\ 
\citep{Udalski91} before performing the aperture photometry. 
The results were checked with 
\textsc{daohphot} profile fitting, where the line-of-sight star was removed
before performing aperture photometry, and the results were found to be the same.
Optimal subtraction 
\citep{Naylor98}, which is well suited for crowded fields was also peformed 
and the results were found to be the same
as using aperture photometry with a mask. 
The normalized Stokes parameters $q$ and $u$ are then determined as well as 
the fractional linear polarization $p$ and position angle $\theta$. The 
errors in the counts are dominated by photon shot noise while the 
theoretical errors in $p$ were estimated by propagation of errors. In 
addition, we calculated the errors in $p$ ($\sigma_p$) using a Monte Carlo 
method, which returned values similar to those estimated from error 
propagation.  Given the large values for $p/\sigma_p$ ($\sim$30) no 
correction for  statistical polarization bias was performed 
\citep{Wardle74}; the bias has the effect of increasing the estimated $p$ 
if the errors on $q$ and $u$ are large (usually due to low signal-to-noise 
data), because $p$ is a positive quantity whereas $q$ and $u$ can be 
positive or negative.

For the TNG data the Stokes $q$ and $u$ parameters of the target and other 
stars were determined and corrected for the instrumental polarization as 
described in \citet{Covino14}. The modelling of the instrumental 
polarization was done using the unpolarized standard stars over a wide 
range of parallactic angles and with PAOLO it has been shown that it can 
be done to better than $\sim$0.2 per cent. The procedure to derive an 
instrumental model implies a correction for the position angle of the 
instrument. After applying the instrumental model we determine the 
instrument-corrected $q$ and $u$ light curves and hence $p$ and 
$\theta$. For the NOT data, the instrumental polarization was determined 
from the zero polarized standard star and was found to be negligible, less 
than 0.1 per cent.

In Table\,\ref{table:results} we give the optical polarization results 
obtained for \target\ during quiescence. In Fig.\,\ref{fig:ppa} we show 
the $R$-band $p$ as a function of $\theta$ for \target\ and other stars in 
the field-of-view. Both $p$ and $\theta$ for \target\ are similar to the 
values measured from the field stars, which cluster around a common value. 
The mean optical VRIz position angle of \target\ is $\sim$6.4$^\circ$.

\subsection{WHT Near-IR}
\label{sec:near-IR}

We observed \target\ with the Long-slit Intermediate Resolution Infrared 
Spectrograph (LIRIS) in imaging polarimetry mode on the 4.2-m William 
Herschel Telescope (WHT) at the Observatorio del Roque de los Muchachos, 
La Palma, Spain. The data were taken on 2015 July 27 \textsc{ut} 22:01 to 
22:40. Exposures were made in a five-point dither pattern, separately in 
the $J$, $H$ and $K_{\rm s}$ filters. The Wollaston prism splits the 
incoming light into four simultaneous images, one at each of the four 
polarization angles; 0$^\circ$, 45$^\circ$, 90$^\circ$ and 135$^\circ$.  
We made use of the achromatic half-wave plate which cancels the relative 
transmission factors of the ordinary and extraordinary rays for each 
Wollaston and saves observing time since camera rotation significantly 
increases overheads. In principle this removes the need to observe a 
polarized standard star as the position angle offset should be zero. The 
conditions were good with a median seeing of 0.6 arcsec.

The data reduction was performed using the \textsc{lirisdr} package 
developed by the LIRIS team in the \textsc{iraf}\footnote{\textsc{iraf} is 
distributed by the National Optical Astronomy Observatory, which is 
operated by the Association of Universities for Research in Astronomy, 
Inc., under cooperative agreement with the National Science Foundation. 
http://iraf.noao.edu/} environment \citep[for details see][]{Alves11}.  
Aperture photometry was then performed on the resulting combined images 
using a fixed aperture ($\sim$2 times the average seeing) and the 
normalized Stokes parameters $q$ and $u$, and $p$ and $\theta$ were 
determined using the equations that apply to half-wave plate data. Errors on $p$ and $\theta$ were computed using a Monte 
Carlo routine that propagates the errors associated with the raw counts at 
each polarization angle. The instrumental polarization is not significant 
and is known to be very small for LIRIS; $<$ 0.1 per cent \citep{Alves11}.  
Given the large values for $p/\sigma_p$ ($>$50) no correction for for 
statistical polarization bias was performed \citep{Wardle74}.

In Table\,\ref{table:results} we give the near-IR polarization results 
obtained for \target\ and in Fig.\,\ref{fig:ppa} we plot the near-IR 
$p$ as a function of $\theta$ for \target\ and other stars in the 
field-of-view. Both $p$ and $\theta$ for \target\ are similar to the 
values measured from the field stars, which cluster around a common value.  
The mean near-IR position angle of \target\  is $\sim$ -6$^\circ$.

\begin{figure}
\centering
\includegraphics[angle=-90,width=7.cm]{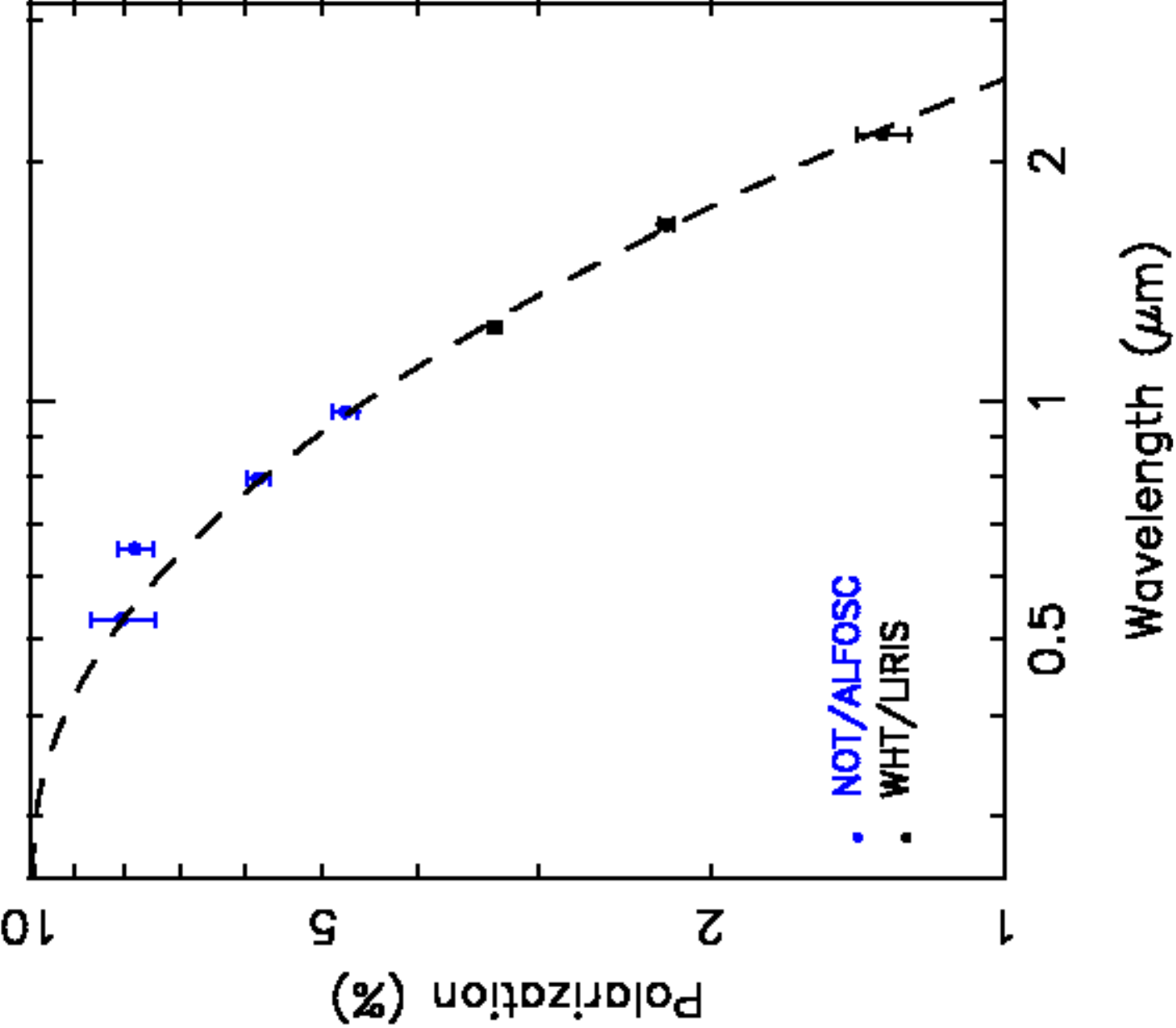}
\caption{
The polarization spectral energy  distribution taken during (and near)
quiescence. The optical NOT (blue filled circles)  were taken on 2016 May 24
($\sim$MJD\,57192.20) and near-IR points (black  filled circles) were taken on
2015 July 27 ($\sim$MJD\,57230.43;  Section\,\ref{sec:near-IR}). The dashed line
shows an  interstellar polarization model  fit to the data.
}
\label{fig:ispol}
\end{figure}

\begin{figure*}
\centering
\includegraphics[angle=-90,width=17cm]{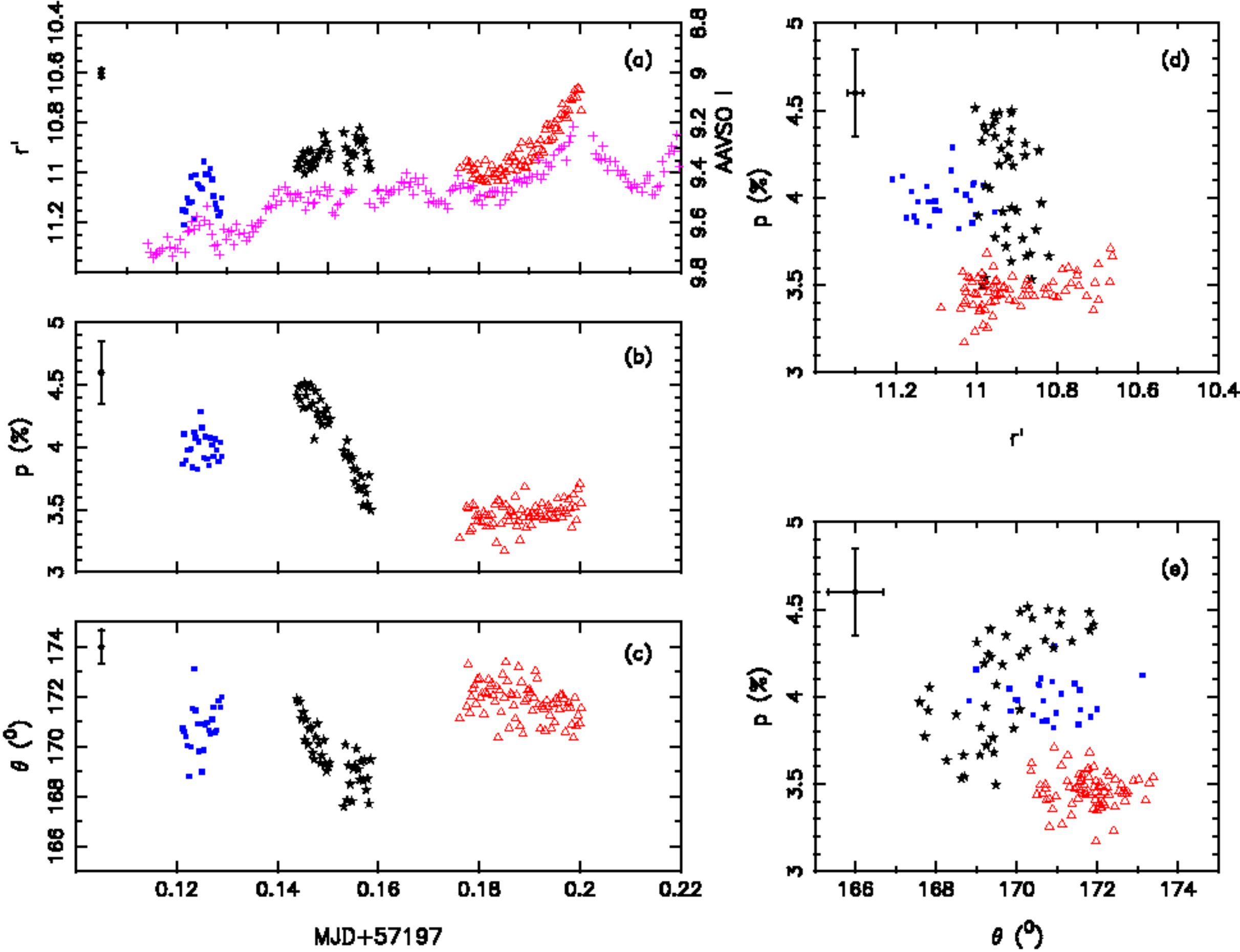}
\caption{ (a) Magnitude, (b) intrinsic polarization and  (c) position angle
light curves. (d) polarization versus magnitude and (e)  polarization versus
position angle plots. In panel (a) the crosses (cyan) are the AAVSO I-band data 
plotted to show the similarity to our 
light curve. The 
polarization data have been  corrected for interstellar dust as
explained in the text.  The different colours refer to the data with 
high  (filled blue squares), low (open red triangles) and variable 
(black filed stars) polarization. The error bar in  each
plot shows the typical uncertainty.
}
\label{fig:pol}
\end{figure*}

\section{RESULTS}

\subsection{Polarization spectral energy distribution}
\label{pol_sed}

During the 2015 outburst of \target\ there were a number of linear polarization 
measurements taken at different times which have very similar $R$-band 
polarization values; $\sim$8 per cent on Jun 17 \citep{Itoh15}, 
7.5$\pm0.1$ on Jun 18 \citep{Blay15} and 7.7$\pm$0.1 per cent on Jun 18 
and Jun 19 \citep{Panopoulou15}, 7.8 per cent on Jun 20 \citep{Tanaka16}. 
All latter measurements have almost identical values, suggesting that it 
is most likely the constant level of the interstellar polarization. We 
also find that the $R$-band polarization level we measure in quiescence 
(Table\,\ref{table:results}) is very similar to these values, which 
further suggests that this polarization level of 7--8 per cent is not due 
to some intrinsic origin present only during outburst. \target\ has a dust 
extinction of $A_{\rm v} = 2.8$--4.4 \citep{Shahbaz03, Hynes09} so 
interstellar optical polarization is expected. A linear relation between 
the maximum optical polarization caused by interstellar dust 
\citep{serket75} implies a polarization level of $< 13$ per cent for 
\target, which is consistent with the observed 7--8 per cent.

In one case, variable optical polarization was reported in \target\ during 
its outburst \citep{Lipunov16} which implies that part of the optical 
polarization is due to an intrinsic, variable component, most likely from 
synchrotron radiation from a jet (see Section\,\ref{sec:disc}). An 
intrinsic component arising from the accretion disc is also likely to be 
present, but this is expected to be very low, at the 1 per cent level 
\citet{}. Therefore, given the high extinction, the polarization of 
\target\ determined when the source is in quiescence, represents the dust 
component along the light-of-sight.

To determine the dust component we fit the optical and near-IR 
polarization measurements with the interstellar dust model.  The wavelength 
dependence of the interstellar polarization (due to aligned dust grains) can 
be described empirically using the Serkowski law \citep{Serkowski74};
\begin{equation}
p(\lambda) = p_{\rm max} \exp[-K\ln^2 (\lambda_{\rm max}/\lambda) ]
\end{equation}

\noindent
where $p(\lambda)$ is the percentage polarization at wavelength $\lambda$, 
$p_{\rm max}$ is the maximum polarization which occurs at wavelength 
$\lambda_{\rm max}$. The curve width parameter $K$ has been found to be 
closely related to $\lambda_{\rm max}$, with the most recent 
determination from observations of 105 stars being K=0.01 + 
1.66$\lambda_{\rm max}$ \citep{Whittet92}. We fit the optical and near-IR 
linear polarization data with the empirical formula of Serkowski and find 
$\lambda_{\rm max}$\,=\,0.30$\pm0.03$\,$\mu$m and $p_{\rm 
max}$\,=\,9.65$\pm$0.36 per cent (see Fig.\,\ref{fig:ispol}). The optical ($\sim 6^\circ$ )
and near-IR ($\sim -6^\circ$ ) position angle is expected to be constant, although there seems to be an offset 
of $\sim 12^\circ$ between them, which is most likely due to a non-zero near-IR 
position angle offset (see Section\,\ref{sec:near-IR}). 
The constant position angle with a steep decrease in polarization towards 
longer wavelengths supports the interstellar origin of the polarization 
observations taken during quiescence, and near quiescence for the near-IR 
data. Since the near-IR data fit the interstellar law so well, we find no 
clear evidence for an additional intrinsic source of polarization in the 
near-IR data which were taken near the end of the outburst.

\subsection{Time resolved polarization light curve of \target}

To determine the intrinsic $q$ and $u$ values of \target\ during outburst, 
we subtract the $q$ and $u$ values obtained from the NOT $R$-band data 
(the $r$ and $R$ central wavelengths are very similar) from the $r'$-band 
$q$ and $u$ time resolved values of \target\ taken during outburst. 
Fig.\,\ref{fig:pol} shows the $r'$-band, $p$ and $\theta$ light curves 
intrinsic to \target. As one can see the polarization is variable, with a 
mean of 3.7 per cent and a variation of from 3.2 to 4.5 per cent, whereas 
the mean position angle is 171$^\circ$ with a variation from 167$^\circ$ 
to 173$^\circ$. There appears to be a rise, then fall in the level of 
polarization while the optical flux is relatively steady, followed by a 
more constant, lower polarization level while the optical flux is rising.

\begin{figure*}
\centering
\hspace*{10mm}
\includegraphics[angle=-90,width=17.cm]{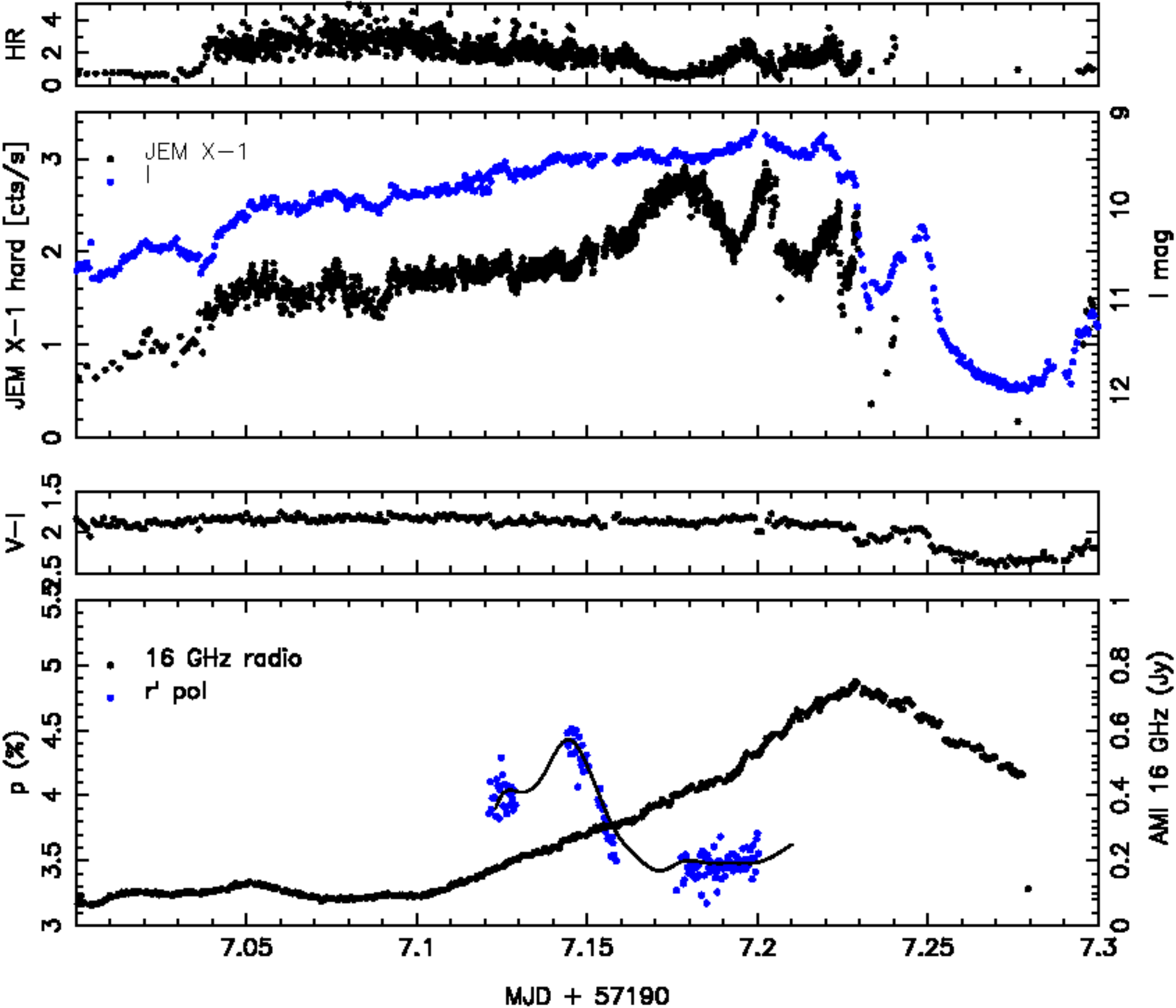}
\caption{
The X-ray, optical and radio light curves near the time of our TNG  observations
on MJD\,57197.  From top to bottom: INTEGRAL JEM-X 10--25/5--10 keV hardness
ratio \citep{Kuulkers15, Kuulkers16}; INTEGRAL JEM-X 5--10\,keV X-ray count rate
(black dots) with the  AAVSO $I$-band light curve (blue dots);  AAVSO
\citep{AAVSO} optical colour; and the AMI-LA 16\,GHz radio data (black dots;
Fender et al. in preparation); with our TNG $r'$-band polarization light curve
(blue dots).
}
\label{fig:multi}
\end{figure*}

\section{DISCUSSION}
\label{sec:disc}

\subsection{Origin of the variable polarization}

In order to constrain the origin of the polarization flare, we compare its 
properties to what is expected from possible sources of polarization. The 
polarization flare seen in 
Fig.\,\ref{fig:pol}b
decayed fairly smoothly from 4.5 to 3.5 per cent in 
approximately 20 minutes, and the mean level of polarization before and 
after the flare differed. Interstellar dust between \target\ and the Earth 
has already been subtracted, so the remaining variable polarization must 
originate within the binary system. Variable (i.e. moving) screens of dust 
or lighter matter \emph{local to} the system, perhaps produced by the 
outburst itself, are unlikely to be the cause of the variable 
polarization. A fast, neutral accretion disc wind is produced during the 
outburst \citep{Munoz16}, but any optical extinction caused by the wind 
would reduce the flux and redden the optical colour, as well as possibly 
change the fractional polarization. During the 20-minute polarization 
flare, the optical flux and colour do not change (Fig.\,\ref{fig:pol}a and 
Fig.\,\ref{fig:multi}), so extinction changes can be ruled out.

Thompson scattering of thermal, unpolarized photons from the accretion 
disc could produce variable polarization, but the timescale of variability 
is expected to be on the orbital period and stronger at bluer wavelengths 
\citep[e.g.][]{Brown78, Cheng88}. This has been observed in some X-ray 
binaries, with a modulation of polarization seen on the orbital period in 
systems in outburst \citep{Michalsky75, Gliozzi98} and quiescence 
\citep{Dolan89}. The orbital period of \target\ is 6.5 days; much longer 
than the timescales of variability we observe, and longer timescale 
polarization studies during the outburst \citep{Lipunov16,Tanaka16} have 
not detected a modulation. It is therefore unlikely that scattering of 
unpolarized light causes the variable polarization.

The bright optical flaring seen throughout the outburst of \target\ 
probably has both thermal (from the accretion disc) and non-thermal 
(synchrotron jet) components \citep[see][and references 
therein]{Gandhi16}. Synchrotron jet emission at optical--near-IR 
wavelengths in X-ray binaries is associated with emission close to the 
base of the jet, and is found to be highly variable from hour down to 
sub-second timescales 
\citep[e.g.][]{Hynes03,Casella10,Gandhi11,Kalamkar16} and polarized, 
usually at a level of a few per cent \citep{Dubus06,Russell08,Shahbaz08, 
Chaty11,Russell14,Baglio14}. Jet synchrotron emission is therefore a 
viable candidate for the variable polarization.

\citet{Rodriguez15} identify and label various flare events during the 
outburst of \target.  Flare XIII reached a flux level of $\sim$40\,Crab, 
with a peak flux lasting $\sim$1.5\,h that was preceded by a $\sim$3\,h 
long, 3\,Crab plateau observed only above 13\,keV. With the exception of 
two flare events, IV and XIII, most of the optical flares show a fast rise 
($\sim$1\,h) similar to the flares observed in X-rays. Flare events IV and 
XIII have slower rise times (about 10 and 4 h, respectively), and are both 
coincident with hard plateaus that precede the X-ray peaks. An optical and 
X-ray plateau is observed before a very bright flare. 
Fig.\,\ref{fig:multi} shows the INTEGRAL JEM-X (5--10 keV) and AAVSO 
$I$-band light curves of \target\ around the time of our TNG polarimetry 
(MJD\,57197.12 to 57197.20). As one can see, a huge X-ray flare 
\citep[labelled flare XIII in][]{Rodriguez15} was observed with JEM-X from 
MJD\,57197.15 to 57197.20, reaching $\sim$ 40 Crab in flux and during 
which the hardness ratio decreases.
The top two panels of Fig.\,\ref{fig:multi} 
show that the X-ray flux and its hardness ratio (black dots)
were highly variable, especially soon after the polarization flare. A bright,
prominent radio (16 GHz) flare was observed with the Arcminute Microkelvin
Imager Large Array (AMI-LA), peaking just a few hours after the optical polarization flare 
(AMI-LA data are from Fender et al. in preparation). 
Note that the AMI-LA does not give polarisation information. 
Concurrently, the optical flux
(bottom panel; data from AAVSO) brightens to a peak flare a few hours after the
polarization flare, and then fades rapidly, with two smooth dips in the light
curve and a change in the colour following the flare. The timing of the
polarization flare does not correspond to any specific feature in either the
X-ray or optical light curves; it occurs during the plateau / slow rise of the
optical and X-ray fluxes before the brightest flares. The polarization flare
does however occur during the initial stages of the rise of the radio flux.

If the optical synchrotron emission originates near the jet base, and the 
bright flare at radio wavelengths is predominantly from emission 
downstream at larger scales further down the jet, then they may be 
associated with the propagation of one single ejection event. This is 
supported by 5 GHz radio data from the e-Multi-Element Radio Linked 
Interferometer Network (eMERLIN; from Fender et al. in preparation; data 
not shown here) which also show a peak in the light curve. The 16\,GHz and 
5\,GHz flares peak at $\sim$2.0\,h and $\sim$\,3.8 h respectively, after 
the $r'$-band polarization flare. We first observe the jet base in 
polarization at optical wavelengths, and then the radio flares, which 
arise from emission along the jet downstream. The three flares describe a 
classically evolving synchrotron flare from an ejection. This strongly 
suggests that a jet launching event occured at the time of the 
polarization flare, and implies that we have discovered the signature of a 
major ejection being launched, from optical polarization.

The 16 GHz radio flare peaks at a flux density of $\sim 0.75$ Jy. The 
optical magnitude at the time of the polarization flare is $r' = 10.85$, 
which corresponds to a de-reddened (assuming $A_v = 4.0$) flux density of 
3.8 Jy. The synchrotron emission may contribute a fraction of the flux, 
and since optically thin synchrotron emission can be up to $\sim 70$ per 
cent polarized, the flux density of the synchrotron could be as low as 
0.25 Jy. If the optical synchrotron flux density is 0.25 -- 3.8 Jy then 
the spectral index from 16 GHz to $r'$-band (at $4.8 \times 10^{14}$ Hz) 
at the flare peak is $\alpha = +0.03 \pm 0.13$ (where $F_{\nu} \propto 
\nu^{\alpha}$), consistent with being a flat spectrum. This is similar to 
what was found for the near-IR and radio flares of GRS 1915+105, in which 
oscillations in $K$-band (2.2 $\mu$m) and 15 GHz radio (2.0 cm) had 
similar timescales, morphologies and flux densities 
\citep{Fender98,Mirabel98}. Continuously launched, hard state jets 
typically have flat or slightly inverted spectra, that are produced by the 
overlapping synchrotron spectra of many individual plasma ejections. Here, 
it appears we are witnessing an individual ejection evolving from optical 
to radio wavelengths, over a frequency range spanning 4.5 orders of 
magnitude.

Miller-Jones et al. (in prep.) present Very Long Baseline Array data from 
the 2015 outburst of \target, showing evidence that the resolved radio jet 
orientation changes over time.  A preliminary analysis suggests that the 
position angle of the jet varied between $\sim$-16$^\circ$ and 
$\sim$+10$^\circ$ E of N. For optically thin synchrotron emission, the 
polarization position angle is parallel to the electric vector and 
perpendicular to the magnetic field vector.  We measure a rather stable 
polarization position angle of $\sim$-9 $^\circ$ E of N from the optical 
data. This implies that the electric field vector near the base of the jet 
in \target\ is on average approximately parallel to the jet axis, implying 
that the magnetic field is orthogonal to the jet axis. This may be due to 
the compression of magnetic field lines in shocks in the flow, resulting 
in a partially ordered transverse field.

\subsection{Comparison to other polarimetry measurements}

\citet{Lipunov16}  also observed  variable linear polarization in  \target\
during its 2015 outburst. Even though they could not determine the absolute
value for the polarization,  they report changes of 4--6 per cent over
timescales  of $\sim$1 h during two different epochs.

During several nights within the outburst, \citet{Tanaka16} obtained 
$R$-band polarimetry of \target\ and several bright stars. They found that 
the $R$-band $p$ and $\theta$ of \target\ and the surrounding field stars 
showed similar levels and direction, respectively. They also determined 
the $V$ to $K_{\rm s}$-band polarization and position angle spectrum and 
concluded that the decrease in polarization towards the infrared with 
constant positon angle, suggests that the polarization is interstellar in 
origin. Our broad-band polarimetry also suggests an interstellar dust 
component dominates the overall polarization spectral energy distribution 
(see Section\,\ref{pol_sed}). However, the fact that we observe a variable 
component \citep[as did][]{Lipunov16}, suggests that there is an 
additional source for the variable polarization, and we argue above that 
this is most likely due to synchrotron emission from a jet.

\citet{Tanaka16} also obtained simultaneous $R$- and $K_{\rm s}$-band 
light curves on MJD\,57193 and 57194 that showed very similar structures; 
a long plateau of slowly rising flux, followed by a rapid decline in flux. 
However, on MJD\,51793 a $K_{\rm s}$-band near-IR flare was observed 
peaking around MJD\,57193.5, $\sim$2\,h before the shoulder of the 
subsequent  rapid decline in flux, with no corresponding enhancement in the 
$R$-band. The observed red colour and short duration of the near-IR flare 
suggest synchrotron emission from a jet. Indeed, the $K_{\rm s}$-band peak 
flux density of the flare reached the same level as the giant radio and 
sub-mm flares observed by RATAN-600 and SMA (Sub Millimeter Array) on 
MJD\,57198.933 and MJD\,57195.55, respectively 
\citep{Trushkin15,Tetarenko15}. As shown in \citet{Tanaka16}, the radio 
flux observed during the giant radio and sub-mm flares extrapolates to the 
$K_{\rm s}$-band peak flux density by assuming a flat spectral index. This 
is similar to the flat spectrum found between our optical polarization 
flare and the following radio flare.

During this near-IR flare reported by \citet{Tanaka16}, at first glance 
the $K_{\rm s}$-band polarization and position angle does not show any 
significant temporal variation, indicating that the near-IR emission is 
not strongly polarized.  However, the errors on the polarization are 
large, and the $K_{\rm s}$-band polarization does in fact change from 
$\sim 2.0$ per cent to $\sim 0.5$ per cent during the near-IR flare, which 
is slightly larger than the change in polarization that we detected (with 
much higher significance) from the TNG data (from 4.5 to 3.5 per cent; see 
Fig.\,\ref{fig:pol}b). The reason why the near-IR polarization change is 
not significant is because the errors are large enough that these 
fluctuations are consistent with the scatter in polarization at other 
times. Therefore one cannot uncover small amplitude 
flares in the polarization. We therefore cannot rule out a small 
polarization flare of amplitude $\sim 1$ per cent being present during the 
near-IR flux flare observed by \citet{Tanaka16} which, if present, would 
be in line with our results. However, the low level of $K_{\rm s}$-band 
polarization suggests that either the synchrotron emission is optically 
thin and the magnetic field is predominantly tangled, or that the jet 
synchrotron emission is in the optically thick regime and the 
magnetic field is highly ordered. Since no 
corresponding $R$-band flare is observed, the optically thick synchrotron 
emission perhaps does not extend up to the $R$-band. The fact that we 
observe a polarized flare in the $r'$-band, while the optical flux 
remained steady, suggests that the majority of the optical flux does not 
originate in the polarized jet component, and that the jet component 
itself is polarized by a stronger amount than observed. It could therefore 
be that the jet synchrotron emission is in the optically thin regime and 
that the spectral break from optically thick to thin lies between $R$ and 
$K_{\rm s}$-band \citep[at a frequency of $\nu_{\rm break} \sim 
1.4-4.7 \times 10^{14}$\,Hz; this is also implied in Fig.\,5 
of][]{Tanaka16}. This frequency for the jet spectral break is consistent 
with that found from data taken during the hard state decay of the 1989 
outburst of \target\ \citep[$\nu_{\rm break} \sim 1.6-2.1 \times 
10^{14}$ Hz;][]{Russell13}.

\begin{figure}
\centering
\includegraphics[angle=-90,width=8.cm]{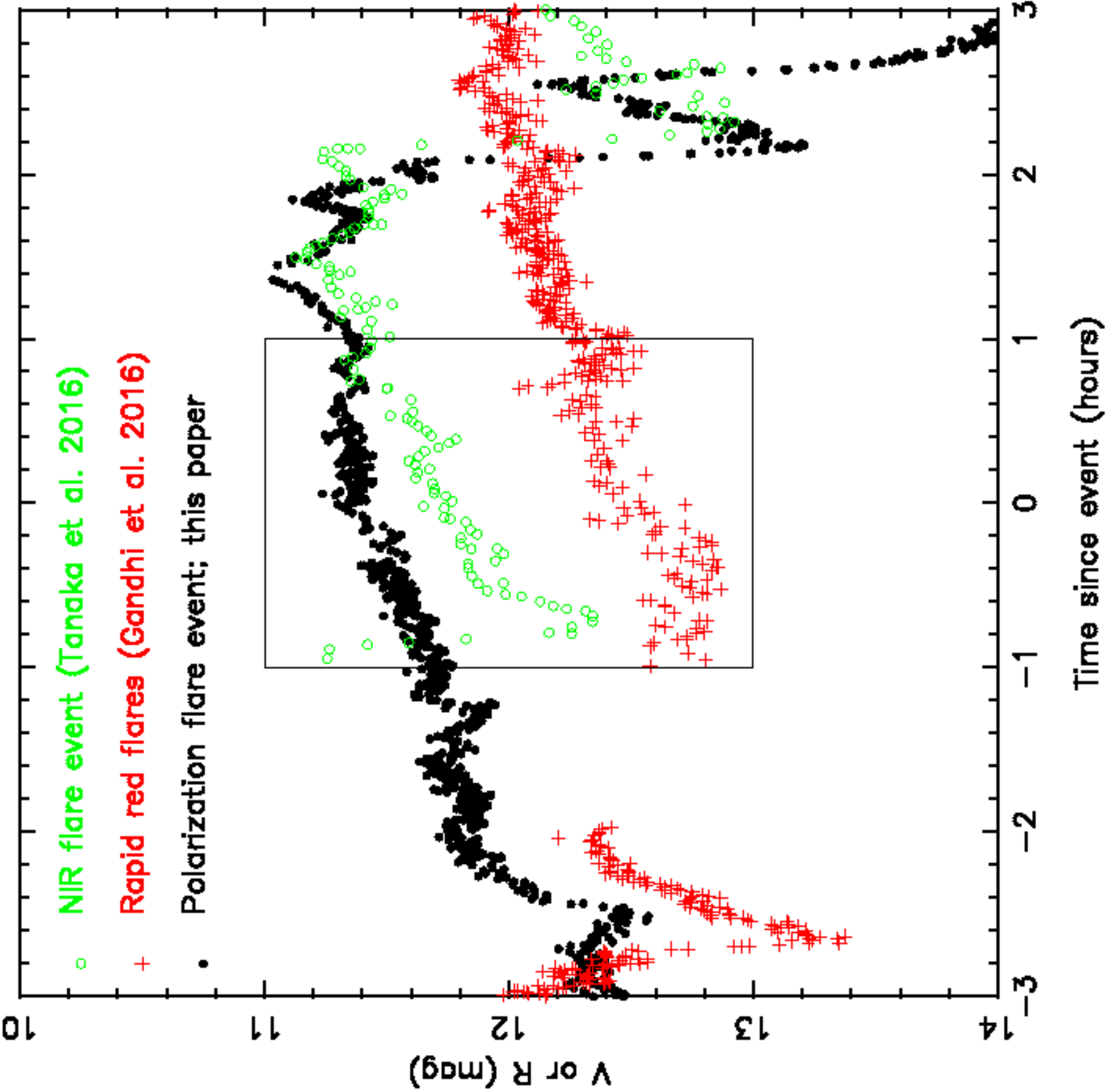}
\caption{
The AAVSO $V$-band lightcurve and the \citet{Tanaka16} $R$-band optical light
curve  (offset for clarity) centred on three  events: the optical polarization
flare reported here (black filled circles; zero is  MJD\,57197.142), the near-IR
flux flare seen by \citet{Tanaka16} (green open circles;  MJD\,57193.543) and
the time of the rapid red flares seen by \citet{Gandhi16} (red  crosses;
MJD\,57199.215). It is clear that all three events occur during a steady,  slow
rise of the flux before some of the brightest flares of the outburst. 
}
\label{fig:overlap}
\end{figure}

\subsection{Repetitive cycles of jet production}

\citet{Kimura16} show that there are significant periods of repetitive 
optical variations during the 2015 outburst of \target. For example, the 
optical light curves taken on MJD\,57197 (see Fig.\,\ref{fig:multi}), 
57193 and 57194 \citep{Tanaka16} show very similar structures; a long 
plateau before a rapid decline in flux. However, on MJD\,51793 a $K_{\rm 
s}$-band flare was observed peaking around MJD\,57193.5, $\sim$2\,h before 
the shoulder of the subsequent rapid decline in flux, with no 
corresponding enhancement in the $R$-band. The near-IR flare suggests at 
least two processes are responsible for the optical--near-IR emission. A 
similar conclusion was reached by \citet{Gandhi16}, who discovered 
remarkable, high amplitude sub-second optical flaring on one date; 26 
June. These rapid flares had a red spectrum and were interpreted as 
originating from optically thin synchrotron emission from the compact jet; 
the same as \citet{Tanaka16} concluded for the origin of the near-IR 
flare. The polarization flare seen with TNG, and the near-IR flare 
reported by \citet{Tanaka16} both occured during the rise of a bright 
optical and 3\,Crab X-ray plateau that precedes a very bright flare (see 
Fig.\,\ref{fig:multi}).

In Fig.\,\ref{fig:overlap} we show the AAVSO and \citet{Tanaka16} optical 
light curves centred on three events: the optical polarization flare 
reported here (black pluses; zero is MJD 57197.142), the near-IR flux 
flare seen by \citet{Tanaka16} (green crosses; MJD 57193.543) and the time 
of the rapid red flares seen by \citet{Gandhi16} (red crosses; MJD 
57199.215). It is clear that all three events occurred during a steady, 
slow rise of the flux (i.e., the plateau) before some of the brightest 
flares of the outburst. In the case of the polarization flare and the 
near-IR flare, the two happened almost exactly 2 hours before a rapid drop 
in the optical flux, in fact the shape of the rise and drop are very 
similar on the two dates. Since all three events are associated with core 
jet activity, we conclude that ejections are produced at this specific 
stage of the flaring cycles of \target\ during its 2015 outburst. In 
addition, some of the brightest radio and mm flares, which trace the large 
scale jet emission, occurred simultaneously with the brightest optical and 
X-ray flares, \citep{Rodriguez15,Munoz16}, further implying that single 
ejection events are being produced during the flare rises. This is 
certainly the case for the optical polarization flare 
(Fig.\,\ref{fig:multi}).

One possible interpretation follows. The 3-Crab X-ray plateau during which 
the polarization flare is detected, could be associated with many smaller 
ejections (a more continuous jet), and would therefore be associated with 
more shocks in the jet ejecta close to the core. Evidence of the launching 
of many small ejections comes from the rapid optically thin flares 
detected also during a plateau stage by \citet{Gandhi16}. The polarization 
flare \citep[and possibly the near-IR flare seen by][]{Tanaka16} could 
signify compression of the magnetic field due to many small shocks in the 
jet during this plateau. This naturally explains the polarization position 
angle observed, which is consistent with a shock compressed magnetic field 
perpendicular to the jet axis, like the high polarization observed in 
downstream shocks in AGN jets, and the high radio polarization in X-ray 
binaries from discrete ejecta/shocks \citep[e.g.][]{Hannikainen00, 
Fender02, Brocksopp07, Curran14}. In this interpretation, the polarization 
flare signifies the birth of a major ejection event caused by an 
intermittent outflow and resulting in a prominent radio flare.
This is not dissimilar
to what has been seen in the past in GRS\,1915+105
\citep[see for example
Fig.\,4 of][]{Klein02}.

\section{CONCLUSIONS}

We find:

\begin{itemize}

\item that the optical to near-IR linear polarization spectrum can be described
by interstellar dust and an intrinsic variable component

\item that the intrinsic optical polarization, detected during the rise of one
of the brightest flares of the outburst, is variable, peaking at 4.5 per cent
and decaying to 3.5 per cent

\item that we favour a synchrotron jet origin to this variable polarization,
with the optical emission originating close to the jet base

\item that the 16GHz and 5GHz flares peak at $\sim$2 h and $\sim$4 h respectively,
after the $r'$-band polarization flare, which is consistent with a classically
evolving synchrotron flare from an ejection.

\end{itemize}

We conclude that we  have witnessed a
plasma ejection evolving over 4.5 orders of magnitude in frequency, from optical
to radio, and that the optical polarization flare represents a jet launching
event; the birth of a major ejection.  For this event we measure a rather
stable  polarization position angle of $\sim$-9 $^\circ$ E of N from the
optical  data. This implies that the electric field vector near the base of the
jet  in \target\ is on average approximately parallel to the jet axis, implying 
that the magnetic field is orthogonal to the jet axis. This may be due to  the
compression of magnetic field lines in shocks in the flow, resulting  in a
partially ordered transverse field. We also find that this ejection occured at a
similar stage in the repetitive cycles of flares, as other indicators of jet
launching found previously: rapid sub-second optical flaring, and excess
near-infared emission.

\section*{ACKNOWLEDGEMENTS}

We thank the referee Prof. Phil Charles for his valuable comments.
We thank James Miller-Jones and Gregory Sivakoff for information related to the
imaging of the  resolved radio jet of \target\ that will appear in Miller-Jones
et al. in preparation.  This article is based on scheduled observations made
with the  William Herschel Telescope, the Telescopio Nazionale Galileo and the
Nordic Optical Telescope operated on the island of La Palma by the Isaac Newton
Group,  the Fundaci\'\o{}n Galileo Galilei of the INAF (Istituto Nazionale di
Astrof\'\i{}sica) and the Nordic Optical Telescope Scientific Association,
respectively,  in the Spanish Observatorio del Roque de los Muchachos of the
Instituto de Astrof\'\i{}sica de Canarias.  This research has been supported by
the Spanish Ministry of Economy and Competitiveness (MINECO) under the grant
AYA\,2013-42627.  
RPF acknowledges support from ERC Advanced Investigator Grant 267607 '4 PI SKY'.
We acknowledge with thanks the variable star observations from
the AAVSO International Database contributed by observers worldwide and used in
this research. Swift/BAT transient monitor results were provided by the
Swift/BAT team.  CR is grateful for the support of STFC Studentships and also
acknowledges the support of Cambridge University. This work has been supported
by ASI grant I/004/11/2.

\footnotesize{

}


\begin{thebibliography}{}

\bibitem[\protect\citeauthoryear{Alves et al.}{2011}]{Alves11} 
Alves F.~O., Acosta-Pulido J.~A., Girart J.~M., Franco G.~A.~P., L{\'o}pez 
R., 2011, AJ, 142, 33 

\bibitem[\protect\citeauthoryear{Barthelmy et al.}{2015}]{Barthelmy15} 
Barthelmy, S. D., D'Ai, A., D'Avanzo, P., Krimm, H. A., Lien, A. Y., 
Marshall, F. E., Maselli, A., Siegel, M. H. 2015, GRB Coordinates Network, 17929, 1

\bibitem[\protect\citeauthoryear{Baglio et al.}{2014}]{Baglio14} 
Baglio M.~C., Mainetti D., D'Avanzo P., Campana S., Covino S., Russell D.~M., Shahbaz T., 2014, A\&A, 572, A99 

\bibitem[\protect\citeauthoryear{Bernardini et al.}{2016}]{Bernardini16}
Bernardini F., Russell D.~M., Shaw A.~W., Lewis F., Charles P.~A., Koljonen K.~I.~I., Lasota J.~P., Casares J., 2016, ApJ, 818, L5

\bibitem[\protect\citeauthoryear{Bjornsson \& Blumenthal}{1982}]{Bjornsson82} 
Bjornsson C.-I., Blumenthal G.~R., 1982, ApJ, 259, 805 

\bibitem[\protect\citeauthoryear{Blandford K{\"o}nigl}{1979}]{Blandford79} 
Blandford R.~D., K{\"o}nigl A., 1979, ApJ, 232, 34 

\bibitem[\protect\citeauthoryear{Blay et al.}{2015}]{Blay15} 
Blay P., Munoz-Darias T., Kajava J., Casares J., Motta S., Telting J., 
2015, ATel, 7678, 1 

\bibitem[\protect\citeauthoryear{Brocksopp et al.}{2007}]{Brocksopp07} 
Brocksopp C., Miller-Jones J.~C.~A., Fender R.~P., Stappers B.~W., 2007, MNRAS, 378, 1111 

\bibitem[\protect\citeauthoryear{Brown, McLean, \& Emslie}{1978}]{Brown78} 
Brown J.~C., McLean I.~S., Emslie A.~G., 1978, A\&A, 68, 415 

\bibitem[\protect\citeauthoryear{Casares, Charles, \& Naylor}{1992}]{Casares92} 
Casares J., Charles P.~A., Naylor T., 1992, Natur, 355, 614 

\bibitem[\protect\citeauthoryear{Casella et al.}{2010}]{Casella10} 
Casella P., et al., 2010, MNRAS, 404, L21 


\bibitem[\protect\citeauthoryear{Charles \& Coe}{2006}]{Charles06} 
Charles P.A.,  Coe, M.J., in Walter Lewin and  Michiel van der Klis, 
eds, Cambridge Astrophysics Series, No. 39, Compact stellar X-ray sources.
Optical, ultraviolet and infrared observations of X-ray binaries. 
Cambridge University Press,  Cambridge, p. 215 

\bibitem[\protect\citeauthoryear{Chaty, Dubus, \& Raichoor}{2011}]{Chaty11} 
Chaty S., Dubus G., Raichoor A., 2011, A\&A, 529, A3 

\bibitem[\protect\citeauthoryear{Cheng et al.}{1988}]{Cheng88} 
Cheng F.~H., Shields G.~A., Lin D.~N.~C., Pringle J.~E., 1988, ApJ, 328, 
223 

\bibitem[\protect\citeauthoryear{Covino et al.}{2014}]{Covino14} 
Covino S., et al., 2014, AN, 335, 117  

\bibitem[\protect\citeauthoryear{Curran et al.}{2014}]{Curran14} 
Curran P.~A., et al., 2014, MNRAS, 437, 3265 

\bibitem[\protect\citeauthoryear{Dhillon et al.}{2007}]{Dhillon07} 
Dhillon V. S. et al., 2007, MNRAS, 378, 825

\bibitem[\protect\citeauthoryear{Dolan}{1984}]{Dolan84}
Dolan J.~F., 1984, A\&A, 138, 1 

\bibitem[\protect\citeauthoryear{Dolan \& Tapia}{1986}]{Dolan86} 
Dolan J.~F., Tapia S., 1986, BAAS, 18, 968 

\bibitem[\protect\citeauthoryear{Dolan \& Tapia}{1989}]{Dolan89} 
Dolan J.~F., Tapia S., 1989, PASP, 101, 1135 

\bibitem[\protect\citeauthoryear{Dubus \& Chaty}{2006}]{Dubus06} 
Dubus G., Chaty S., 2006, A\&A, 458, 591 

\bibitem[\protect\citeauthoryear{Fender \& Pooley}{1998}]{Fender98} 
Fender R.~P., Pooley G.~G., 1998, MNRAS, 300, 573 

\bibitem[\protect\citeauthoryear{Fender et al.}{2002}]{Fender02} 
Fender R.~P., Rayner D., McCormick D.~G., Muxlow T.~W.~B., Pooley G.~G., Sault R.~J., Spencer R.~E., 
2002, MNRAS. 336, 39

\bibitem[\protect\citeauthoryear{Ferrigno et al.}{2015}]{Ferrigno15} 
Ferrigno C., et al., 2015, ATel, 7662, 1 

\bibitem[\protect\citeauthoryear{Gandhi et al.}{2011}]{Gandhi11}
Gandhi P., et al., 2011, ApJ, 740, L13 

\bibitem[\protect\citeauthoryear{Gandhi et al.}{2016}]{Gandhi16} 
Gandhi P., et al., 2016, MNRAS, 459, 554 

\bibitem[\protect\citeauthoryear{Garner et al.}{2015}]{Garner15} 
Garner A., et al., 2015, ATel, 7663, 1 

\bibitem[\protect\citeauthoryear{Giro et al.}{2003}]{Giro03} 
Giro E., Bonoli C., Leone F., Molinari E., Pernechele C., Zacchei A., 2003, 
SPIE, 4843, 456 

\bibitem[\protect\citeauthoryear{Gliozzi et al.}{1998}]{Gliozzi98} 
Gliozzi M., Bodo G., Ghisellini G., Scaltriti F., Trussoni E., 1998, A\&A, 337, L39 

\bibitem[\protect\citeauthoryear{Hannikainen et al.}{2000}]{Hannikainen00} 
Hannikainen D.~C., Hunstead R.~W., Campbell-Wilson D., Wu K., McKay D.~J., Smits D.~P., Sault R.~J., 2000, 
ApJ, 540, 521 

\bibitem[\protect\citeauthoryear{Hynes et al.}{2003}]{Hynes03} 
Hynes R.~I., et al., 2003, MNRAS, 345, 292 

\bibitem[\protect\citeauthoryear{Hynes et al.}{2009}]{Hynes09} 
Hynes R.~I., Bradley C.~K., Rupen M., Gallo E., Fender R.~P., Casares J., 
Zurita C., 2009, MNRAS, 399, 2239 

\bibitem[\protect\citeauthoryear{Hynes et al.}{2004}]{Hynes04} 
Hynes R.~I., et al., 2004, ApJ, 611, L125 

\bibitem[\protect\citeauthoryear{Itoh et al.}{2015}]{Itoh15} 
Itoh R., et al., 2015, ATel, 7709, 1 

\bibitem[\protect\citeauthoryear{Kafka}{2016}]{AAVSO} 
Kafka, S., 2016, Observations from the AAVSO International Database, http://www.aavso.org

\bibitem[\protect\citeauthoryear{Kalamkar et al.}{2016}]{Kalamkar16} 
Kalamkar M., Casella P., Uttley P., O'Brien K., Russell D., Maccarone T., van der Klis M., Vincentelli F., 2016, MNRAS, in press (arXiv:1510.08907)

\bibitem[\protect\citeauthoryear{Kimura et al.}{2016}]{Kimura16} 
Kimura, M., et al., 2016, Natur, 529, 54

\bibitem[\protect\citeauthoryear{Kitamoto et al.}{1989}]{Kitamoto89} 
Kitamoto S., Tsunemi H., Miyamoto S., Yamashita K., Mizobuchi S., 1989, Natur, 342, 518 

\bibitem[\protect\citeauthoryear{Klein-Wolt et al.}{2002}]{Klein02}
Klein-Wolt M., Fender R.~P., Pooley G.~G., Belloni T., Migliari S., Morgan E.~H., van der Klis M., 2002, MNRAS, 331, 745 


\bibitem[\protect\citeauthoryear{Kong et al.}{2002}]{Kong02} 
Kong A.~K.~H., McClintock J.~E., Garcia M.~R., Murray S.~S., Barret D., 2002, ApJ, 570, 277 

\bibitem[\protect\citeauthoryear{Krimm et al.}{2013}]{Krimm13} 
Krimm H.~A., et al., 2013, ApJS, 209, 14 

\bibitem[\protect\citeauthoryear{Kuulkers \& Ferrigno}{2016}]{Kuulkers16} 
Kuulkers E., amp, Ferrigno C., 2016, ATel, 8512,  

\bibitem[\protect\citeauthoryear{Kuulkers}{2015}]{Kuulkers15} 
Kuulkers E., 2015, ATel, 7758,  

\bibitem[\protect\citeauthoryear{Kuulkers et al.}{2015}]{Kuulkers15} 
Kuulkers E., Motta S., Kajava J., Homan J., Fender R., Jonker P., 2015, ATel, 7647, 1 

\bibitem[\protect\citeauthoryear{Landi Degl'Innocenti, Bagnulo, \& Fossati}{2007}]{Landi07} 
Landi Degl'Innocenti E., Bagnulo S., Fossati L., 2007, ASPC, 364, 495


\bibitem[\protect\citeauthoryear{Lipunov et al.}{2016}]{Lipunov16} 
Lipunov V.~M., et al., 2016, arXiv, arXiv:1608.02764 

\bibitem[\protect\citeauthoryear{Makino et al.}{1989}]{Makino89} 
Makino F., Wagner R.~M., Starrfield S., Buie M.~W., Bond H.~E., Johnson J., 
Harrison T., Gehrz R.~D., 1989, IAUC, 4786, 1 

\bibitem[\protect\citeauthoryear{Mart{\'{\i}}, Luque-Escamilla, \& Garc{\'{\i}}a-Hern{\'a}ndez}{2016}]{Marti16} 
Mart{\'{\i}} J., Luque-Escamilla P.~L., Garc{\'{\i}}a-Hern{\'a}ndez M.~T., 2016, A\&A, 586, A58 

\bibitem[\protect\citeauthoryear{Michalsky, Swedlund, \& Stokes}{1975}]{Michalsky75} 
Michalsky J.~J., Swedlund J.~B., Stokes R.~A., 1975, ApJ, 198, L101

\bibitem[\protect\citeauthoryear{Miller-Jones et al.}{2009}]{Miller-Jones09} 
Miller-Jones J.~C.~A., Jonker P.~G., Dhawan V., Brisken W., Rupen M.~P., Nelemans G., Gallo E., 2009, ApJ, 706, 
L230 

\bibitem[\protect\citeauthoryear{Mirabel et al.}{1998}]{Mirabel98} 
Mirabel I.~F., Dhawan V., Chaty S., Rodriguez L.~F., Marti J., Robinson C.~R., Swank J., Geballe T., 1998, A\&A, 330, L9 

\bibitem[\protect\citeauthoryear{Mooley et al.}{2015}]{Mooley15} 
Mooley K., Fender R., Anderson G., Staley T., Kuulkers E., Rumsey C., 2015, 
ATel, 7658, 1 

\bibitem[\protect\citeauthoryear{Motta et al.}{2015a}]{Motta15a} 
Motta S., Beardmore A., Oates S., Sanna N.~P.~M.~K.~A., Kuulkers E., Kajava 
J., Sanchez-Fernanedz C., 2015a, ATel, 7665, 1 

\bibitem[\protect\citeauthoryear{Motta et al.}{2015b}]{Motta15b} 
Motta S., Beardmore A., Oates S., Sanna N.~P.~M.~K.~A., Kuulkers E., Kajava 
J., Sanchez-Fernanedz C., 2015b, ATel, 7666, 1 

\bibitem[\protect\citeauthoryear{Mu{\~n}oz-Darias et al.}{2016}]{Munoz16} 
Mu{\~n}oz-Darias T., et al., 2016, Natur, 534, 75 

\bibitem[\protect\citeauthoryear{Natalucci et al.}{2015}]{Natalucci15} 
Natalucci L., Fiocchi M., 
Bazzano A., Ubertini P., Roques J.-P., Jourdain E., 2015, ApJ, 813, L21 

\bibitem[\protect\citeauthoryear{Naylor}{1998}]{Naylor98} 
Naylor T., 1998, MNRAS, 296, 339 

\bibitem[\protect\citeauthoryear{Negoro et al.}{2015}]{Negoro15} 
Negoro H., et al., 2015, ATel, 7646, 1 

\bibitem[\protect\citeauthoryear{Oliva}{1997}]{Oliva97} 
Oliva E., 1997, A\&AS, 123,  

\bibitem[\protect\citeauthoryear{Panopoulou, Reig, \& Blinov}{2015}]{Panopoulou15} 
Panopoulou G., Reig P., Blinov D., 2015, ATel, 7674, 1 

\bibitem[\protect\citeauthoryear{Rodriguez et al.}{2015}]{Rodriguez15} 
Rodriguez J., et al., 2015, A\&A, 581, L9 

\bibitem[\protect\citeauthoryear{Russell \& Fender}{2008}]{Russell08} 
Russell D.~M., Fender R.~P., 2008, MNRAS, 387, 713 

\bibitem[\protect\citeauthoryear{Russell et al.}{2013}]{Russell13}
Russell D.~M., et al., 2013, MNRAS, 429, 815

\bibitem[\protect\citeauthoryear{Russell \& Shahbaz}{2014}]{Russell14} 
Russell D.~M., Shahbaz T., 2014, MNRAS, 438, 2083 

\bibitem[\protect\citeauthoryear{Rybicki \& Lightman}{1979}]{Rybicki79} 
Rybicki G. B., Lightman A. P. 1979, in Radiative Processes in Astrophysics, New York, Wiley

\bibitem[\protect\citeauthoryear{Schultz}{2000}]{Schultz00} 
Schultz J., 2000, A\&A, 364, 587 

\bibitem[\protect\citeauthoryear{Serkowski}{1974}]{Serkowski74} 
Serkowski K., 1974, in  Methods  of  Experimental  Physics,  Vol.  12A,  ed.  N.
Carleton (New York: Academic), 361

\bibitem[\protect\citeauthoryear{Serkowski, Mathewson, \& Ford}{1975}]{serket75} Serkowski K., Mathewson D.~S., Ford V.~L., 1975, ApJ, 196, 261 


\bibitem[\protect\citeauthoryear{Spruit}{2014}]{Spruit14} 
Spruit H,, 2014, in 
Gonz{\'a}lez Mart{\'{\i}}nez-Pa{\'{\i}}s I., Shahbaz T., 
Casares Vel{\'a}zquez J., eds, 
XXI Canary Islands Winter School of Astrophysics, 
Accretion discs. Cambridge University Press, Cambridge, p. 1

\bibitem[\protect\citeauthoryear{Shahbaz et al.}{1996}]{Shahbaz96} 
Shahbaz T., Bandyopadhyay R., Charles P.~A., Naylor T., 1996, MNRAS, 282, 977 

\bibitem[\protect\citeauthoryear{Shahbaz et al.}{1994}]{Shahbaz94}
Shahbaz T., Ringwald F.~A., Bunn J.~C.,  Naylor T., Charles P.~A., Casares J., 1994, MNRAS, 271, L10 

\bibitem[\protect\citeauthoryear{Shahbaz et al.}{2003}]{Shahbaz03} 
Shahbaz T., Dhillon V.~S., Marsh T.~R., Zurita C., Haswell C.~A., Charles P.~A., Hynes R.~I., Casares J., 2003, 
MNRAS, 346, 1116 

\bibitem[\protect\citeauthoryear{Shahbaz et al.}{2008}]{Shahbaz08} 
Shahbaz T., Fender R.~P., Watson C.~A., O'Brien K., 2008, ApJ, 672, 510 

\bibitem[\protect\citeauthoryear{Tanaka et al.}{2016}]{Tanaka16} 
Tanaka Y.~T., et al., 2016, arXiv, arXiv:1601.01312 
 

\bibitem[\protect\citeauthoryear{Tetarenko et al.}{2015}]{Tetarenko15} 
Tetarenko A., Sivakoff G.~R., Young K.,  Wouterloot J.~G.~A., 
Miller-Jones J.~C., 2015, ATel, 7708, 1 

\bibitem[\protect\citeauthoryear{Trushkin, Nizhelskij, \& Tsybulev}{2015}]{Trushkin15} 
Trushkin S.~A., Nizhelskij N.~A., Tsybulev P.~G., 2015, ATel, 7716, 1 


\bibitem[\protect\citeauthoryear{Tsubono et  al.}{2015}]{Tsubono15} 
Tsubono K., Aoki T., Asuma K., Daishido T., Kida S., Nakajima H., 
Niinuma K., Takefuji K., 2015, ATel, 7701, 1 


\bibitem[\protect\citeauthoryear{Udalski \& Kaluzny}{1991}]{Udalski91} 
Udalski A., Kaluzny J., 1991, PASP, 103, 198 

\bibitem[\protect\citeauthoryear{van Paradijs \& McClintock}{1994}]{vanParadijs94} 
van Paradijs J., McClintock J.~E., 1994, A\&A, 290,  

\bibitem[\protect\citeauthoryear{Wagner et al.}{1994}]{Wagner94} 
Wagner R.~M., Starrfield S.~G., Hjellming R.~M., Howell S.~B., Kreidl T.~J., 1994, ApJ, 429, L25 

\bibitem[\protect\citeauthoryear{Wardle \& Kronberg}{1974}]{Wardle74} 
Wardle J.~F.~C., Kronberg P.~P., 1974, ApJ, 194, 249 

\bibitem[\protect\citeauthoryear{Whittet et al.}{1992}]{Whittet92} 
Whittet D.~C.~B., Martin P.~G., Hough J.~H., Rouse M.~F., Bailey J.~A., Axon
D.~J., 1992, ApJ, 386, 562 


\end{thebibliography}
\end{document}